\documentclass[12pt,letterpaper]{article}
\usepackage{amsmath,amssymb,epsfig}
\numberwithin{equation}{section}
\newcommand{\be}{\begin{equation}}
\newcommand{\bea}{\begin{eqnarray}}
\newcommand{\eea}{\end{eqnarray}}
\newcommand{\ba}{\begin{array}}
\newcommand{\ea}{\end{array}}
\newcommand{\ee}{\end{equation}}

\expandafter\ifx\csname mathbbm\endcsname\relax

\else

\fi \textheight 22cm \textwidth 15cm \topmargin 1mm \oddsidemargin
5mm \evensidemargin 5mm

\def\t{\tau}
\def\la{\lambda}
\def\n{\nu}
\def\p{\phi}
\def\v{\varphi}
\def\r{\rho}

\begin{document}

\begin{titlepage}
 \hfill
 \vbox{
    \halign{#\hfil         \cr
           hep-th/0411087 \cr
           IPM/P-2004/066 \cr
           } 
      }  
 \vspace*{20mm}
 \begin{center}
 {\Large {\bf On Isolated Conformal Fixed Points and
 Noncritical String Theory}\\ }

 \vspace*{15mm} \vspace*{1mm} {Mohsen Alishahiha, Ahmad Ghodsi and Amir E. Mosaffa}

 \vspace*{1cm}

 {\it  Institute for Studies in Theoretical
 Physics and Mathematics (IPM)\\
 P.O. Box 19395-5531, Tehran, Iran\\}

 \vspace*{1cm}
 \end{center}

 \begin{abstract}
 We search for the gravity description of unidentified field theories at their
 conformal fixed points by studying the low energy effective action of six
 dimensional noncritical
 string theory. We find constant dilaton solutions by solving both the equations of motion
 and BPS equations. Our solutions include a free parameter provided
 by a stack of
 uncharged space filling branes. We find several $AdS_p\times S^q$ solutions with
 constant radii for $AdS_p$ and $S^q$. The curvature of the solutions are of
 the order of the string scale.

 \end{abstract}

 \vspace{2cm}
 \begin{center}
 \it{Dedicated to Farhad Ardalan on the occasion of his 65th birthday}
 \end{center}
 \end{titlepage}

 \section{Introduction}
The conjectured duality of string and gauge theories found its
first explicit realization in the context of AdS/CFT
correspondence
\cite{{Maldacena:1997re},{Gubser:1998bc},{Witten:1998qj}}. This
correspondence is based upon the analysis of D3 branes in IIB
string theory and its low energy effective SUGRA in the limit
where the branes are replaced by their RR fluxes. The original
form of AdS/CFT and its modifications relate critical IIB string
theories and four dimensional conformal gauge theories in the
complementary regimes of the string sigma model and gauge
couplings.

This strong/weak nature of the duality enables one to
obtain lots of information about some strongly coupled gauge
theories by studying the corresponding low energy supergravities.
However, this ideology breaks down when the coupling of the CFT
cannot be varied arbitrarily. Examples of such CFT's can be found
in certain $\cal{N}$$=1$ gauge theories \cite{Seiberg:1994pq}.

This and other considerations have motivated the study of
noncritical strings
 in search for the gravity duals of gauge
theories that have isolated fixed points in their renormalization
group flow \cite{Polyakov:1998ju}\footnote{ For further studies in this direction
see \cite{Klebanov:1998yy}-\cite{Ghoroku:1999bk}.}. Such theories are suggested to be
dual to noncritical
strings with target spaces of the form $AdS_p\times S^q$.\footnote{This
scenario has mainly studied for the noncritical string in dimensions
less that 10. Noncritical string for $d>10$ has also been studied, for example,
in \cite{Maloney:2002rr} where it was shown that in comparison with the $d<10$ case
it exhibits new feature, such as a dilaton potential with nontrivial
minima at small cosmological constant and $d$-dimensional string coupling.}

By studying the one loop $\beta$ function of noncritical sigma
models, it has been shown that the corresponding low energy
effective actions admit solutions of this form with constant radii
for the $AdS_p$ and $S^q$ spaces. In fact these solutions have curvatures
of the order of the string scale. Therefore the one loop approximation becomes
inadequate and higher loops must be taken into account. One
expects that in general this will destroy the $AdS\times S$
structure of the solutions but it is believed that this is not
the case and the corrections will simply modify the corresponding
radii of the solution.

This idea was reinforced in \cite{Polyakov:2004br} where several simple
noncritical $AdS_p\times S^q$ sigma models were suggested by
demanding a conformally invariant  worldsheet theory.
Although the all order invariance was not proved, several arguments
were presented in favor of that.
It was also shown that the sigma
models should necessarily have $\kappa$ symmetry to allow for
renormalizability and thus are, as in the critical case,
completely integrable.

It was also mentioned \cite{Polyakov:2004br} that in order to get a gravity solution in
the form of $AdS_p\times S^q$ with a free parameter, one may add flavors
to the corresponding gauge theory.

Following this idea, the authors of
\cite{Klebanov:2004ya} considered 3 brane solutions in presence of
space filling $D5/\bar{D}5$ brane pairs in six dimensional
noncritical string theory and were able to find an $AdS_5\times
S^1$ solution with fixed radii for $AdS_5$ and $S^1$. In this construction
the space filling $D5/\bar{D}5$ branes play the role of the flavors
in the corresponding gauge theory. Therefore getting an $AdS_5$
solution
could be expected because the
solution must correspond to $\cal{N}$$=1$ superconformal gauge
theory which necessarily includes matter.

By arguments regarding stability of the solution
and also the open
string tachyons on the $D5$-branes, the free parameter was restricted to a
certain range which was suggested to be identical to the Seiberg
conformal window of the corresponding gauge theory. It was also
speculated that Seiberg's electric-magnetic duality is related to
T-duality on the $S^1$.

The supergravity solutions for noncritical string in $d>1$ dimensions
have also been obtained in \cite{Kuperstein:2004yk} where several
solutions in the form of $AdS_p\times S^q$ were presented. In this
paper the authors have considered cases where only RR or NS forms are nonzero.
As a result they have not been able to find
the solution of \cite{Klebanov:2004ya} where the flavors
play an essential role.

It is the aim of this paper to further study  more general brane solutions
in six dimensional string theory. With the goal of a better understanding
of unidentified field theories with conformal fixed points, we mainly focus
on solutions with a constant dilaton.

We keep the $D/\bar{D}$ pairs in the setup to have a free parameter in our
solutions.
In particular it also
enables us to find solutions like $AdS_3\times S^3$ which are not
possible in the absence of flavors. Apart from solving the
standard equations of motion, we insist on finding solutions from
the BPS equations, whenever possible, and thus meet the necessary
requirement for supersymmetry of solutions.

The organization of the paper is as follows. In section 2 we
introduce the general setup of the problem and give a brief review
of some of the known results.
In section 3 we study gravity solutions in the presence of $N_f$
space filling uncharged D5-brane. In the subsequent two sections
we generalize to the cases where either an RR or an NS charge is turned on.
In section 6 we give a second derivation of the solutions by using
BPS equations. We conclude by discussions.

\section{Basic Setup}

In this section we shall fix our notations and the basic setup to be
used in the next sections. Our starting point is  six
dimensional noncritical string theory in the presence of different
fluxes. In general the theory could contain $N_f$ space filling
uncharged D5-branes, NS three form $H_3$ and nonzero RR $k$-forms
$F_k$. For this general background the bosonic part of the effective action
is given by
 \bea
 S&=& \int dx^6 \sqrt{-g}
 \bigg[e^{-2\phi}\bigg(R+4(\partial\phi)^2+c\bigg)-2N_fe^{-\phi}\bigg]
 \cr && \cr
 &-&\frac{1}{2}\int dx^6\sqrt{-g}\bigg[\frac{e^{-2\phi}}{3!}H_3^2+\sum_{k}
 \frac{1}{k!}F_k^2\bigg]\;,
 \label{action}
 \eea
 where the action is written in the string frame. We work in units
 where $\alpha'=1$ such that the cosmological constant term which
 is the characteristic of the non-critical theory will be
 \be
 c=\frac{10-d}{\alpha'}=4\;.
 \ee
 We will keep this constant in our calculations to trace the effect
 of non-criticality.

 Let us consider the following ansatz for the metric in the string frame
 \be
 ds^2=d\t^2+e^{2\la(\t)}dx_n^2+e^{2\n(\t)}d\Omega_k^2\;.
 \label{ansatz}
 \ee
 where $\t$ is the radial coordinate, $dx_n^2$ is an
 $n$ dimensional flat space which is associated with the world
 volume of a $D_{n-1}$ brane and $d\Omega_k^2$ is a $k$ sphere and $n+k+1=6$.
 For the RR form we take the magnetic ansatz $F_k\sim\omega_k$
 with $\omega_k$ being the volume form of $S_k$. This will describe
 a $(4-k)$ D-brane. Similarly for the NS form we take
 $H_3\sim\omega_3$ which will give an NS1 brane. It is clear that
 the NS three form can contribute only when $k=3$.

 In the following we will only study one nonzero charge, NS or RR, at a time.
 We will thus be considering $D_p$-branes with $p=1,2,3,4$ and
 NS1 brane.
 The relations obtained below will be in fact the
 ones found in \cite{Kuperstein:2004yk} which are modified
 due to the $D/\bar{D}$ pairs.

 Plugging (\ref{ansatz}) into the action (\ref{action}) and doing the possible
 simplifications one arrives at the following form for the action
 \bea
 S&=&\int
 d\rho\bigg[-n(\la')^2-k(\n')^2+(\v')^2+ce^{-2\v}+k(k-1)e^{-2\n-2\v}\bigg]
 \cr && \cr
 &-&\int
 d\rho\bigg[N^2e^{n\la-k\n-\v}+Q^2e^{-2k\n-2\v}+2N_fe^{\frac{1}{2}(n\la+k\n)-\frac{3}{2}
 \v}\bigg]\;,
 \label{action2}
 \eea
 where prime denotes the derivatives with respect to $\rho$ with
 $d\t=-e^{\v}d\rho$. $N$ and $Q$ are proportional to the number of D$_{k-4}$ and
 NS1 branes respectively and $\v$ is defined as
 \be
 \v=2\p-n\la-k\n\;.
 \ee

 The equations of motion derived from this action read
 \bea
 \la''_\r&-&\frac{1}{2}N^2e^{2n\la-2\p}-\frac{1}{2}N_fe^{2n\la+2k\n-3\p}=0\;,
 \cr && \cr
 \n''_\r&-&(k-1)e^{2n\la+2(k-1)\n-4\p}+\frac{1}{2}N^2e^{2n\la-2\p}
 +Q^2e^{2n\la-4\p}\cr && \cr
 &-&\frac{1}{2}N_fe^{2n\la+2k\n-3\p}=0\;,
 \cr && \cr
 \p''_\r&+&\frac{c}{2}e^{2n\la+2k\n-4\p}-\frac{1}{4}(n+1-k)N^2e^{2n\la-2\p}
 +\frac{1}{2}(k-1)Q^2e^{2n\la-4\p}\cr && \cr
 &-&\frac{1}{4}(n+k+3)N_fe^{2n\la+2k\n-3\p}=0\;.
 \label{eom}
 \eea
 To simplify the equations we have written them in terms of $\p$
 instead of $\v$.
 There is a fourth equation, known as the zero energy constraint,
 which is in fact the equation of motion for $g_{\t\t}$\footnote{Note that
 the derivatives in (\ref{eom}) and (\ref{cons}) are with respect to $\r$
 and $\t$ respectively.}
 \bea
 n(\la'_\t)^2&+&k(\n'_\t)^2-(\v'_\t)^2+c+k(k-1)e^{-2\n}\cr && \cr
 &-&N^2e^{n\la-k\n+\v}-Q^2e^{-2k\n}
 -2N_fe^{\frac{1}{2}(n\la+k\n+\v)}=0\;.
 \label{cons}
 \eea

 There is a method for finding first order
 differential equations of motion instead of the second order
 ones obtained above. These equations are known as
 BPS equations of motion and the method is based on defining a
 superpotential $W$ from which the equations are derived.
 Satisfying the BPS equations is a
 necessary but not sufficient condition for the solution to be
 supersymmetric. The method works as follows:

 Consider an action of the form
 \be
 S=\int d\r\bigg(-\frac{1}{2}G_{ab}(f){f^a}'{f^b}'-V(f)\bigg)\;.
 \ee
 If one can find a superpotential $W(f)$ such that
 \be
 V(f)=\frac{1}{8}G^{ab}\partial_aW\partial_bW\;
 \label{potsup}
 \ee
 then the BPS equations of motion are written as
 \be
 {f^a}'=\frac{1}{2}G^{ab}\partial_bW\;.
 \ee
 It is easy to see that the effective action (\ref{action2}) can be recast
 into the above form. Doing so, we find that
 \be
 G_{\la\la}=2n\;,\;\;\;\;\;G_{\n\n}=2k\;,\;\;\;\;\;G_{\v\v}=-2\;,
 \ee
 and the potential reads
 \be
 V=-(c+k(k-1)e^{-2\n})e^{-2\v}+N^2e^{n\la-k\n-\v}
 +Q^2e^{-2k\n-2\v}+2N_fe^{\frac{1}{2}(n\la+k\n-3\v)}\;.
 \label{pot}
 \ee
 The relation (\ref{potsup}) will take the following form
 \be
 \frac{1}{n}(\partial_\la W)^2+ \frac{1}{k} (\partial_\v W)^2-(\partial_\v
 W)^2=16V\;.
 \label{BPScon}
 \ee
 and finally the BPS equations are written as
 \be
 \la'_\r=\frac{1}{4n}\partial_\la W\;,\;\;\;\;\;\v'_\r=\frac{1}{4k}\partial_\v
 W\;,\;\;\;\;\;\v'_\r=-\frac{1}{4}\partial_\v
 W\;.
 \ee
 In order to see how the above method works, we briefly review some
 of the known results obtained by the BPS equations.

 Let us consider the case where $N=Q=N_f=0$ in which the potential (\ref{pot})
 will have an over all factor of $e^{-2\v}$. Thus one may take the following
 ansatz for the superpotential
 \be
 W=4e^{-\v}w(\la,\n)\;.
 \ee
 We consider two distinct cases of $k=0,1$. For $k=0$, the
 relation (\ref{BPScon}) is solved by $w(\la)=\pm\sqrt{c}$. With
 this choice, the superpotential will be flat in the $\la$
 direction and a free parameter will appear in the solution,
 $\la_0$. This parameter, however, has no physical significance
 because it can be absorbed by a rescaling of the $n=5$
 dimensional space. The BPS equation for $\p$ is solved by
 $\p=\pm\frac{\sqrt{c}}{2}\t+\p_0$. Of course here the free parameter,
 $\p_0$, has a physical significance which determines the asymptotic
 behavior of the string coupling. The resulting solution will
 be a six dimensional flat space with linear dilaton.

 There is a second choice for $W$
 namely $w(\la)=\pm\sqrt{c}\cosh(\sqrt{5}\la)$. The BPS equations
 are solved by
 \be
 e^{\la}=\bigg[\tanh\bigg(\frac{1}{2}\sqrt{c}\t\bigg)\bigg]^{1/\sqrt{5}}\;,\;\;\;\;\;
 e^{2\p}=\frac{1}{ a}\frac{[\tanh(\frac{1}{2}\sqrt{c}\t)]^{\sqrt{5}}}
 {\sinh(\sqrt{c}\t)}\;,
 \ee
 where $a$ is a constant. The solution has a naked singularity at $\t=0$
 and for $\t\rightarrow\infty$ it reduces to flat space
 with a linear dilaton.

 For $k=1$, $w=\sqrt{c}$ is still a solution, though in this case $W$ is
 flat both in $\la$ and $\n$ directions and the latter
 parameterizes the solutions by the radius of $S^1$. The dilaton
 is found as $\p=\frac{\sqrt{c}}{2}\t+\p_0$ and the resulting geometry
 will be a cylinder.

 As a second solution for $W$ one may take
 \be
 w(\la,\n)=\sqrt{c}\cosh\bigg(\frac{m\lambda
 +n\nu}{\sqrt{\frac{m^2}{4}+n^2}}\bigg)\;
 \ee
 where $m$ and $n$ are two real numbers. For $m=0$ one finds
 \be
 \la=\la_0\;,\;\;\;\;\;e^{\n}=\tanh\bigg(\frac{1}{2}\sqrt{c}\t\bigg)\;,\;\;\;\;\;
 e^{2\p}=\frac{1}{2a}\frac{1}{\cosh^2(\frac{1}{2}\sqrt{c}\t)}\;,
 \ee
 and $a$ is a constant. This solution is known as the ``cigar''
 background. For general $m$ and $n$ one gets
 \be
 e^{\la}=\bigg[\coth\bigg(\frac{1}{2}\sqrt{c}\t\bigg)\bigg]^
 {\frac{\frac{m}{4}}{\sqrt{\frac{m^2}{4}+n^2}}}\;,\;\;\;\;\;
 e^{\n}=\bigg[\coth\bigg(\frac{1}{2}\sqrt{c}\t\bigg)\bigg]^
 {\frac{n}{\sqrt{\frac{m^2}{4}+n^2}}}\;.
 \ee
 Requiring regularity at $\t=0$ one needs to have
 $\frac{m}{ n}=-\frac{3}{4}$. The dilaton is identical to the one for the cigar
 solution.

 Finally, it is interesting to note that in the absence of $N_f,N$
 and $Q$, the action (\ref{action2}) is invariant under the
 transformation $\n\rightarrow-\n$ and $\p\rightarrow\p-\n$. This
 is nothing but T-duality by which one can generate
 new solutions. As an example, applying these transformations to
 the cigar solution one arrives at a new background known as the
 ``trumpet'' solution \cite{Kuperstein:2004yk}.

 The inclusion of of $D/\bar{D}$ pairs can result in serious
 modifications in the problem. For example in \cite{Klebanov:2004ya} it
 was shown that in presence of these pairs a solution of the form
 $AdS_5\times S^1$ is obtained in six dimensions whereas in their absence, as shown
 by \cite{Kuperstein:2004yk}, the $AdS_5\times S^1$ and $AdS_3\times S^3$
 solutions are not possible.

 In the following sections we will further study the effects of
 these space filling brane pairs with and without nonzero $N$ or $Q$.
 Since we are looking for those backgrounds which correspond to field theories at
 their conformal
 fixed point, we will mainly be interested in solutions with a constant
 dilaton. Therefore the third equation in (\ref{eom}) requires that
 \be
 2c-(n+1-k)N^2e^{-2k\n+2\p}+2(k-1)Q^2e^{-2k\n}-(n+k+3)N_fe^{\p}=0\;.
 \label{condil}
 \ee
 This relation will be our starting point for finding
 solutions in the following sections.

 \section{\large{Solutions with $N_f$ space filling uncharged branes}}
 In this section we shall study our first example where we will turn on
 $N_f$ space filling uncharged branes and set all other charges to zero.
 In this case by making use of the equation (\ref{condil}) one can find the
 dilaton which is
 \be
 e^{\p}=\frac{2c}{(n+k+3)N_f}=\frac{1}{N_f}\;.
 \label{dil1}
 \ee
 Looking at the the potential (\ref{pot}) we realize that the two cases $k=0,1$ are
 special because the curvature contribution from $S^k$
 vanishes for them. We shall study these two cases separately.
 For other values of $k$ we could not solve the equations.


 \subsection*{$\bullet\;\;\; k=0$}

 When $k=0$, the only equation one should solve is the one for
 $\la$. It is useful to rewrite the second derivatives in the equations of motion
 with respect to $\t$ instead of $\r$\footnote{From now on all the derivatives
 are written with respect to $\t$ unless stated.}.
 In this notation, taking into account that $\v'=-5\la'$, one arrives at the following
 equation for $\la$
 \be
 \la''+5\la'^2=\frac{N_f}{2}e^{\p}\;.
 \ee
 Here $\p$ is given by (\ref{dil1}). This equation can be solved for
 $\lambda$ and the final solution is
 \be
 \la=\frac{\t}{\sqrt{10}}+\la_0\;.
 \label{ads61}
 \ee
 Therefore the background geometry is given by
 \be
 ds^2=d\tau^2+e^{2\tau/\sqrt{10}}dx_{5}^2\;,
 \ee
 which is an $AdS_6$ background with $R_{AdS}^2=10$. Note that in the above
 metric we have rescaled the coordinates $\vec{x}$ by $e^{2\lambda_0}$.

 It is worth noting that six dimensional string theory supports another
 $AdS_6$ solution which has a nonzero RR five form, no
 space filling uncharged D5-branes and no transverse sphere and
 is obtained as \cite{Kuperstein:2004yk}
 \be
 e^{\p}=\frac{2}{\sqrt{3}N}\;,\;\;\;\;\;\;\;\;\;\;\;\;\;\;\;R_{AdS}^2=\frac{15}{2}\;.
 \label{ads6sonn}
 \ee
 We will see in section 4 that a third $AdS_6$ solution
 with $k=0$ can be obtained when both nonzero RR five form and
 uncharged $D5$ branes are present.

 \subsection*{$\bullet\;\; k=1$}

 In this case it is convenient to introduce two new variables $x$ and $y$ as the following
 \be
 x=\la-\n\;,\;\;\;\;\;\;\;\;\;\;\;\;\;\;\;y=4\la+\n\;.
 \ee
 In terms of these new variables the equations of motion are
 \bea
 y''+y'^2&=&\frac{5}{2}\;,\cr && \cr
 x''+x'y'&=&0\;.
 \label{k=1xy}
 \eea
 It is easy to solve the above equation for $y$ leading to
 \be
 y=\sqrt{\frac{5}{2}}\t+y_0\;.
 \label{k=1y}
 \ee
 One can then proceed to find $x$  from the second equation. The simplest solution can be
 found by setting $x'=0$ which together with the solution for $y$ will give the following
 solution for $\lambda$ and $\nu$
 \be
 \la=\frac{\t}{\sqrt{10}}+\la_0\;,\;\;\;\;\;\;\;\;\;\;\;
 \;\;\;\;\n=\frac{\t}{\sqrt{10}}+\n_0\;.
 \ee
 The resulting geometry is thus again $AdS_6$ with $R_{AdS}^2=10$ but this
 time one of the coordinates is compact, i.e.
 \be
 ds^2=d\t^2+e^{2\t/\sqrt{10}}(dx_4^2+d\theta^2)\;.
 \label{ads6}
 \ee
 Here we have rescaled $\vec{x}$ by $e^{2\lambda_0}$ and set $\n_0=0$.

 It is also possible to find a more general solution for this case which is
 \bea
 \la&=&-\frac{\t}{\sqrt{10}}+\frac{2}{5}\log(e^{\frac{5\t}{\sqrt{10}}}-c_0
 )+\la_0\cr &&\cr
 \n&=&-\frac{\t}{\sqrt{10}}+\log(e^{\frac{5\t}{\sqrt{10}}}+c_0)-
 \frac{3}{5}\log(e^{\frac{5\t}{\sqrt{10}}}-c_0)+\n_0\;.
 \label{nonex}
 \eea
 This solution has an additional free parameter in comparison with the previous one
 and reduces to that for $c_0=0$. Alternatively, (\ref{ads6})
 is the $\tau\rightarrow \infty$ limit of this solution.
 For $x'\neq0$ we were not able to solve the equations.

 \section{Solutions with nonzero $N$ and $N_f$}
 In this section we shall study cases with a nonzero RR form in the presence of
 $N_f$ space filling uncharged branes. The constant dilaton condition leads
 to the following equation
 \be
 2c-(n+1-k)N^2e^{-2k\n+2\p}-(n+k+3)N_fe^{\p}=0\;.
 \label{condil1}
 \ee
 We recognize two different possibilities; if $n+1-k\neq0$ then $\n$
 necessarily has to be a constant but for $n+1-k=0\; (k=3)$, $\n$ may or
 may not be a constant. We shall consider these cases separately.
 From the above equation one can see that for $N_f=0$ a constant
 dilaton solution is not possible for $k=3$ but with a nonzero
 $N_f$ this limitation is removed.

 \subsection*{$\bullet\;\; k=0$}
 In this case the
 equation (\ref{condil1}) can be easily solved for the dilaton
 \be
 e^{\p}=\frac{2}{ N_f}\frac{1}{1+\sqrt{1+3{N^2/ N_f^2}}}\;.
 \ee
 The other parameter can be obtained by solving the equation for $\la$.
 Doing so we find an $AdS_6$ solution
 with radius
 \be
 R_{AdS}^2=5\frac{(1+\sqrt{1+3{N^2/ N_f^2}})^2}{1+2\frac{N^2}{
 N_f^2}+\sqrt{{1+3{N^2/ N_f^2}}}}\;.
 \ee
 This is our first example in which the $AdS$ radius has $N$ and
 $N_f$ dependence. Of course it
 depends only on the ratio of these numbers $N/N_f$. As a
 result this solution represents an
 $AdS_6$ solution with one parameter. It is also interesting to note
 that both $N_f\rightarrow 0$ and
 $N\rightarrow 0$ limits are smooth and the resulting solutions are those
 presented in (\ref{ads6sonn}) and (\ref{ads61}) respectively.
 \subsection*{$\bullet \;\; k=1$}
 This case has been studied in \cite{Klebanov:2004ya} and, for completeness,
 we present its final result.
 The solution turns out to $AdS_5\times S^1$ with the following radii
 for $AdS_5$ and $S^1$ spaces
 \be
 R_{AdS}^2=6,\;\;\;\;\;\;\;\;R_{S^1}^2=\frac{2N^2}{ 3N_f^2}\;.
 \ee
 The Dilaton is also given by
 \be
 e^{\p}=\frac{2}{3N_f}.
 \ee
 It has been argued that in order to get a stable solution one needs to
 restrict $N/N_f$ to values $(\sim1)$ which correspond to the conformal window
 of the dual theory.

 \subsection*{$\bullet \;\; k=2$}
 The constant dilaton condition (\ref{condil1}) in this case,
 is solved by
 \be
 e^\p=\frac{2x-1}{x}\frac{2}{5N_f}\;,
 \ee
 where $x\equiv e^{2\n}=R_{S^2}^2$ and is found from the
 equation for $\n$
 \be
 x^4+2x^3-\frac{1}{5}(\frac{N}{N_f})^2(2x-1)^2=0\;.
 \ee
 The equation for $\la$ is easily solved and results in an $AdS_4\times S^2$ background
 with
 \be
 R_{AdS}^2=\frac{15x}{4x+3}\;.
 \ee
 It is worth noting that since $x$ is a function of $N/N_f$ the $AdS$ radius is also
 a function of that and therefore a one parameter $AdS_4\times S^2$
 solution is obtained.

 \subsection*{$\bullet \;\; k=3$}
 We first find the constant dilaton from
 (\ref{condil1})
 \be
 e^\p=\frac{1}{ N_f}\;.
 \ee
 The remaining equations of motion for $\n$ and $\la$ will read
 \bea
 \la''+2\la'^2+3\la'\n'&=&\frac{1}{2}\frac{N^2}{ N_f^2}e^{-6\n}+\frac{1}{2}\;,\cr
 && \cr
 \n''+3\n'^2+2\la'\n'&=&2e^{-2\n}-\frac{1}{2}\frac{N^2}{
 N_f^2}e^{-6\n}+\frac{1}{2}\;.
 \eea
 As we have mentioned in section 2, in this case one can choose $\n$ to be
 either a constant or not.
 If we take it to be a constant, it is found from the second
 equation above
 \be
 x^3+4x^2-\frac{N^2}{ N_f^2}=0\;,
 \ee
 where now $x\equiv e^{2\n}=R_{S^3}^2$. This relation shows that $x$ is
 a function of $N/N_f$. It is now easy to solve the equation for
 $\la$ and for the final solution we an $AdS_3\times S^3$ with
 \be
 R_{AdS}^2=\frac{2x}{x+2}\;.
 \ee
 Therefore, as in the previous cases, we find a one parameter
 solution. We note that
 similar to the $AdS_5\times S^1$ case, this solution exists only if
 the space filling uncharged D-brane are present which means that the $N_f\rightarrow
 0$ limit is not smooth. For non constant $\n$ we were not able to
 solve the equations.

 \section{Solutions with nonzero $N_f$ and NS two form }
 Let us now study the gravity solution when we have self-dual NS1 brane.
 This corresponds to the situation when we have a nonzero NSNS two for,
 $B_{\mu\nu}$. We consider the case with nonzero $B_{\mu\nu}$ in the present of
 $N_f$ space filling uncharged brane and still look for a solution with
 constant dilaton. As stated before, such solutions are only relevant
 when $k=3$. The constant dilaton condition reads
 \be
 2c=8N_fe^\p-4Q^2e^{-6\n}\;.
 \ee
 This relation implies that $\n$ also has to be a constant. The
 equation for $\n$ will read
 \be
 x^3+4x^2-\frac{3}{2}Q^2=0\;,
 \ee
 where again $x\equiv e^{2\n}=R_{S^3}^2$ and is a function of $Q$.
 The dilaton is also found as
 \be
 e^\p=\frac{x+1}{x}\frac{4}{3N_f}\;.
 \ee
 Once we solve the equation for $\la$ we find that the solution
 will be $AdS_3\times S^3$ with
 \be
 R_{AdS}^2=\frac{3x}{1+x}\;.
 \ee
 We note that here again we find a one
 parameter $AdS_3\times S^3$ solution. This parameter is now given by
 $Q$. We note also that the $N_f\rightarrow 0$ limit is not
 smooth showing that this solution exists only in the presence of
 space filling uncharged branes.

 \section{Constant dilaton solutions from BPS equations}
 In the previous sections we obtained gravity solutions of six dimensional noncritical
 string in the form of $AdS_{n+1}\times S^{5-n}$ for $n=2,3,4,5$ using second
 order differential equations which are the equations of motion. As we mentioned
 before there is a method for finding solutions using first order differential
 equation known as BPS equations.

 We note however that not all the solutions of the second order
 equations can necessarily be obtained form the first order
 equations. In fact satisfying the
 BPS equations is a necessary condition for the solution to be supersymmetric.

 In this section we shall show that all our solutions, except one,
 can indeed be obtained from the
 BPS equations and therefore have a good chance to be supersymmetric.
 The exception is the (\ref{nonex}) solution which is very similar
 to non extremal brane backgrounds with $c_0$ playing the role of
 the horizon radius. It is therefore natural for such a solution
 not to be supersymmetric and thus one cannot obtain it from BPS
 equations.

 Our calculations are in parallel lines with those of
 \cite{Kuperstein:2004yk}. Knowing that we are looking for a constant dilaton
 solution we can find the good variables for the superpotential in
 terms of which the calculations find a simple form.
 It is important to note that the solutions we are looking for do not
 require the superpotential to be independent of
 the dilaton and in fact this dependence is necessary
 because otherwise the constant dilaton would appear as a free
 parameter in the solution. Instead, we look for superpotentials that have
 an extremum at a certain value for the dilaton and thus the
 constant value for the dilaton is fixed.

 Let us first study the solutions of section 3 for the case of $k=1$. The good
 variables for this case are
 \be
 x=\la-\n\;,\;\;\;\;\;\;\;\;\;\;\;\;\;\;\;y=4\la+\v+\n\;,
 \;\;\;\;\;\;\;\;\;\;\;\;\;\;\;z=4\la+5\v+\n\;.
 \ee
 Note that $y=2\p$ and the superpotential we are looking for must have an
 extremum at $y=y_0=-2\ln{N_f}$. We also know from the solution
 that $x$ has to be a constant as well. The field $z$ has been chosen in
 a way that the metric $G^{ab}$ defined in (\ref{potsup}) is
 diagonal. In terms of our new variables the
 potential (\ref{pot}) takes the following form
 \be
 V=e^{-\frac{z}{2}}(-ce^{\frac{y}{2}}+2N_fe^y)\;.
 \ee
 Therefore we will consider the following ansatz for the superpotential
 \be
 W(x,y,z)=e^{-\frac{z}{4}}w(y)\;.
 \ee
 Now putting everything together the relation (\ref{potsup}) becomes
 \be
 -\frac{5}{4}w(y)^2+4\partial_yw(y)^2=16(-ce^{\frac{y}{2}}+2N_fe^y)\;.
 \ee
 It is easy to check that $V$ has an extremum at
 $y_0=-2\ln{N_f}$ which, together with the above equation, tells us
 that $\partial_yw$ also has to vanish at $y_0$. Therefore the BPS
 equations will become
 \be
 x'=0\;,\;\;\;\;\;\;\;\;\;\;\;\;\;\;\;y'|_{y=y_0}=0\;,
 \;\;\;\;\;\;\;\;\;\;\;\;\;\;\;z'|_{y=y_0}=-\frac{20}{\sqrt{10}}\;.
 \ee
 This means that
 \be
 \la'|_{y=y_0}=\n'|_{y=y_0}=\frac{1}{\sqrt{10}}\;,
 \ee
 which obviously produces the solution we were looking for.

 Next we consider the remaining solutions which have an additional
 constant namely $\n=\n_0$.
 We therefore, following \cite{Kuperstein:2004yk}, parameterize the problem
 in terms of $\nu$ and the
 following two variables
 \be
 x=\frac{1}{ n-1}(\la+\v)\;,\;\;\;\;\;\;\;\;\;\;\;\;\;\;\;
 y=\frac{1}{ n-1}(n\la+\v)\;,
 \ee
 where the overall numerical factors are chosen for convenience.
 The potential (\ref{pot}) is written as
 \be
 V=e^{-2nx}V(y,\n)\;,
 \ee
 where
 \be
 V(y,\n)=-ce^{2y}-k(k-1)e^{2y-2\n}+N^2e^{(n+1)y-k\n}
 +Q^2e^{2y-2k\n}+2N_fe^{\frac{3+n}{2}y+\frac{k}{2}\n}\;.
 \label{vxy}
 \ee
 We thus choose the following ansatz for the superpotential
 \be
 W(x,y,\n)=e^{-nx}w(y,\n)\;,
 \ee
 and the relation (\ref{potsup}) becomes
 \be
 -\frac{n}{ n-1}w^2+\frac{1}{ k}\partial_\n w^2+\frac{1}{ n-1}\partial_y
 w^2=16V(y,\n)\,.
 \label{wvxy}
 \ee
 Extremizing $V(y,\n)$ will give us the constant values of $y_0$
 and $\n_0$
 \be
 \partial_yV(y,\n)|_{y=y_0,\n=\n_0}=\partial_{\nu}V(y,\n)|_{y=y_0,\n=\n_0}=0\;.
 \ee
 The relation (\ref{wvxy}) tells us that $\partial_\n w$ and $\partial_y
 w$ vanish at $\n_0$ and $y_0$ and that $w_0^2\equiv w^2(y_0,\n_0)
 =16\frac{1-n}{ n}V(y_0,\n_0)$. Finally, the BPS equation for $\la$ is
 \be
 \la'=\bigg(\frac{V(y_0,\n_o)}{ n(1-n)}\bigg)^{1/2}e^{-y_0}\;.
 \label{lbps}
 \ee
 It is now straight forward to check that all the solutions we
 have obtained so far can be obtained by the above formulation.

\section{Conclusions}
 In this paper we have studied six dimensional noncritical string
 theory by looking at its low energy effective supergravity
 action. We looked for NS and D brane solutions of the equations of motion
 with one nonzero charge at a time. Our motivation was looking for
 backgrounds which can be considered as the gravity duals of
 conformal field theories. We therefore focused on constant dilaton
 solutions.

 This problem has some features which are
 characteristics of noncritical theories. In the absence of branes,
 the equations never yield solutions with a constant dilaton. The
 famous examples are the linear dilaton and cigar solutions. Once
 branes are included, under certain circumstances, constant
 dilaton solutions are found. These solutions have the general
 form of $AdS_p\times S^q$. The situation is completely the
 opposite in critical theories where in the absence of branes
 the dilaton is a constant while the brane
 solutions generically yield a non constant dilaton.
 A famous counter example of course is the $D3$ brane of IIB.

 Things are changed considerably when we add a new ingredient,
 uncharged space filling branes, to the noncritical theory.
 Firstly, one no longer needs nonzero NS or RR charges, or
 equivalently additional branes, to find a constant dilaton
 solution. We have given examples of this in section 3.
 Secondly, some of the brane solutions with constant dilaton which
 were impossible previously can now be obtained. Examples are
 $AdS_5\times S^1$ of \cite{Klebanov:2004ya} and
 $AdS_3\times S^3$ obtained in section 4 of the present work.
 Finally, the previously obtained $AdS_p\times S^q$ solutions are
 now just modified by a new constant value for the dilaton and new
 radii for the two $AdS_p$ and $S^q$ spaces. Examples of these
 cases are also given in section 4. We summarize all these solutions in table 1.
\begin{table}[htbp]
\begin{center}
        \begin {tabular}{|c||c|c|c|}
\hline
  &$N_f\neq0$ & $N\neq0$& $N,N_f\neq0$\\
\hline $k=0$&$AdS_6$ &
$AdS_6$ & $AdS_6$
\\ \hline $k=1$    &$AdS_6$    & -   &
$AdS_5\times S^1$ \\
\hline $k=2$&   &$AdS_4\times S^2$ &$AdS_4\times
S^2$
\\ \hline $k=3$ & &-   &$AdS_3\times S^3$ \\
\hline
\end{tabular}
\end{center}
\caption{$AdS$ solutions in six dimensional noncritical string theory }
\end{table}

 The $AdS_p\times S^q$ solutions we have found all share the
 general feature that the curvature of the $AdS$ space and the
 radius of the transverse sphere are constant. This is expected
 and is another characteristic of noncritical theories. It is in
 fact the property that makes them useful for studying
 field theories with isolated conformal fixed points. The constant
 curvatures of our solutions are all of order one in the string
 scale which makes higher $\alpha'$ corrections necessary. As
 stated before, these corrections are not supposed to change the
 general form of the solutions.

 One may also find $AdS$ black hole solutions in this theory. For example 
 the $AdS_5\times S^1$ black hole solution is obtained as
 \be
 ds^2=\frac{R_{AdS}^2}{u^2}\frac{du^2}{1-\frac{u_0^4}{u^4}}
 -\frac{u^2}{R_{AdS}^2}(1-\frac{u_0^4}{u^4})dt^2
 +u^2 dx_i^2+R_{S^1}^2d\theta^2\,,\label{badsmf}
 \ee
 with the same radii as in the extremal case. 
 
 This solution can be used to study for example a confining gauge theory
 in three dimension. Since this solution has a free parameter namely $N/N_f$,
 one might suspect that it can be useful for having
 control over the glueball and KK masses of the confining theory.
 Four dimensional confining theory has recently been studied in
 \cite{Kuperstein:2004yf} using noncritical $AdS_6$ black hole solution.
 It would be interesting to see how this black hole solution
 and the corresponding confining theory are modified once we
 use the $AdS_6$ solutions of sections 3 and 4 to construct
 the associated black hole solutions.

 We have also given a second derivation of most of our results by
 solving the first order BPS equations. This is a necessary
 condition for the supersymmetry of solutions. The superpotential
 defined for this purpose has an extremum at the point of the
 solutions. Perturbations around these points can in principle
 determine the flows of the equations which asymptote in the one
 end to our solutions. To determine the other end one should
 presumably solve the equations numerically.

 One would also like to know the form of the correspondence for
 the backgrounds under discussion. In order to do this one should
 determine the expansion parameters of the corresponding theories. Consider
 for example the $AdS_5\times S^1$ solution. To validate the
 string tree level approximation $N_f$ has to be very large
 $(e^\p\sim1/N_f)$. This can be achieved in a more familiar form if we define the 't Hooft
 coupling of the field theory as in the conventional AdS/CFT correspondence
 by $\la=g_sN=N/N_f$. Therefore the weak string coupling limit translates into the usual
 't Hooft or large $N$ limit. By considering open string tachyons on
 the D5 branes, it was argued in \cite{Klebanov:2004ya}
 that $N/N_f\sim1$. The sigma model expansion, on the other hand,
 is governed by a coupling which is of order of unity. As a result, the correspondence
 seems to be between classical noncritical string theory on this background and
 a large $N$ field theory at its conformal fixed point where the 't Hooft coupling is
 of order of unity.

 For the backgrounds we have found in section 4, the correspondence
 seems to hold in a similar way. Namely one would expect that $N/N_f\sim1$ still holds,
 and therefore the 't Hooft couplings would still be of order of unity and also the string
 coupling should become very small in the large $N$ limit.

 One may ask what gravity could tell us about the degrees of freedom of the
 ``dual gauge'' theory. To estimate the degrees of freedom one may look at the
 gravity action which in all our cases turns out to be
 \be
 \int \sqrt{g}e^{-2\phi}{\cal R}\sim N_f^2f\bigg(\frac{N}{N_f}\bigg)\;,
 \ee
 where the function $f(x)$ can be determined in each case.
 For example in the $AdS_5\times S^1$ case it is given by $f(x)=x$.
 In general for $x\sim1$ one obtains $f(x)\sim1$. This suggests
 that the degrees of freedom of the dual gauge theory goes like $N^2$.
 Here we have used the fact that
 the 't Hooft coupling is of order of one
 \footnote{We would like to thank J. Maldacena for a comment on this point.}.

\section*{Acknowledgments}
We would like to thank J. Maldacena and M. H. Sarmadi for useful discussions. This work is supported in part by Iranian
TWAS chapter based at ISMO.

\end{document}